# Deep 10 and 18 μm Imaging of the HR 4796A Circumstellar Disk: Transient Dust Particles & Tentative Evidence for a Brightness Asymmetry [1]

C. M. Telesco [2], R. S. Fisher [2], R. K. Piña [2], R. F. Knacke [3],
S. F. Dermott [2], M. C. Wyatt [2], K. Grogan [2], E. K. Holmes [2], A. M. Ghez [4], L. Prato [4],
L. W. Hartmann [5], and R. Jayawardhana [5]

[1] Based on observations at the W.M. Keck Observatory, which is operated as a scientific partnership between Caltech, the University of California, and NASA, and made possible by the generous financial support of the W.M. Keck Foundation.

[2] Department of Astronomy, University of Florida, Gainesville, FL 32611; telesco,fisher,rpina,dermott,grogan,holmes,wyatt@astro.ufl.edu

[3] School of Science, Pennsylvania State University at Erie, Station Road, Erie, PA 16563; rfk2@psu.edu

[4] Department of Astronomy, University of California, Los Angeles, CA 90095; ghez@athena.astro.ucla.edu, lprato@minerva.astro.ucla.edu

[5] Harvard-Smithsonian Center for Astrophysics, 60 Garden Street, Cambridge, MA 02138; hartmann@fuor.harvard.edu, rayjay@shakti.harvard.edu

## ABSTRACT

We present new 10.8 and 18.2 µm images of HR 4796A, a young A0 V star that was recently discovered to have a spectacular, nearly edge-on, circumstellar disk prominent at ~20 µm (Jayawardhana et al. 1998; Koerner et al. 1998). These new images, obtained with OSCIR at Keck II, show that the disk's size at 10 µm is comparable to its size at 18 µm. Therefore, the 18 µm-emitting dust may also emit some, or all, of the 10 µm radiation. Using these multi-wavelength images, we determine a "characteristic" diameter of 2-3 µm for the mid-infrared–emitting dust particles if they are spherical and composed of astronomical silicates. Particles this small are expected to be blown out of the system by radiation pressure in a few hundred years, and therefore these particles are unlikely to be promordial. Rather, as inferred in a companion paper (Wyatt et al. 2000), they are probably products of collisions that dominate both the creation and destruction of dust in the HR 4796A disk. Dynamical modeling of the disk (Wyatt et al. 2000) indicates that the disk surface density is relatively sharply peaked near 70 AU, which agrees with the mean annular radius deduced by Schneider et al. (1999) from their NICMOS images. Interior to 70 AU, the model density drops steeply by a factor of two between 70 and 60 AU, falling to zero by 45 AU, which corresponds to the edge of the previously discovered central hole; in the context of the dynamical models, this "soft" edge for the central hole results from the fact that the dust particle orbits are non-circular. The optical depth of mid-infrared–emitting dust in the hole is ~3% of the optical depth in the disk, and the hole is therefore relatively very empty. We present evidence (~1.8σ significance) for a brightness asymmetry that may result from the presence of the hole and the gravitational perturbation of the disk particle orbits by the low-mass stellar companion or a planet. This "pericenter glow," which must still be confirmed, results from a very small (a few AU) shift of the disk's center of symmetry relative to the central star HR 4796A; one side of the inner boundary of the annulus is shifted towards HR 4796A, thereby becoming warmer and more infrared-emitting. The possible detection of pericenter glow implies that the detection of even complex dynamical effects of planets on disks is within reach.

# 1. INTRODUCTION

Planets form in dusty circumstellar disks, and so the exploration of circumstellar disks contributes fundamentally to a broader understanding of the origin and evolution of our own and other planetary systems. Vega-type disks, often called debris disks, are those that surround main sequence stars (see the review by Backman & Paresce 1993). Unlike the disks surrounding pre-main-sequence stars, the dust masses of debris disks are generally much smaller than $0.01 M_\odot$, which is roughly the minimum mass required for the formation of a planetary system like our own (see, e.g., Beckwith & Sargent 1993; Strom et al. 1993). The lower mass of debris disks conforms with the expectation that disks disperse with time as the material accretes on to the stars, forms planets, and/or is blown out of the systems (e.g., Strom et al. 1993). The elucidation of these evolutionary processes is a key challenge of observational astronomy.

High-angular-resolution, multi-wavelength imaging of circumstellar disks can provide unique insight into these processes, since a disk's detailed structure should bear traces of its history and of any developing or mature planets. Until recently, the disk surrounding the A5 star β Pictoris (Smith & Terrile 1984; Telesco et al. 1988; Lagage & Pantin 1994; Kalas & Jewitt 1995) was the only debris disk that was spatially well-resolved at optical or infrared wavelengths. However, many new results have emerged over the past few years including: possible optical imaging of a disk around the binary star BD+31°643 (Kalas & Jewitt 1997); submillimeter imaging of impressive structure in several debris disks, including the class archetype Vega (Holland et al. 1998; Greaves et al. 1998); and ~20 μm imaging of the nearly edge-on disk around the A0V star HR 4796A (Jayawardhana et al. 1998 [J98]; Koerner et al. 1998 [K98]), results that were recently complemented with NICMOS images revealing what appears to be a sharply defined ring (Schneider et al. 1999).

Here we present new 10.8 and 18.2 μm images, obtained at the Keck II telescope, of the HR 4796A debris disk. Jura (1991) first drew attention to HR 4796, pointing out that its infrared spectral energy distribution resembles that of β Pic but with twice the IR excess ($L_{IR}/L_*$). The IRAS data showed no discernable emission from dust hotter than ~110 K, implying that there is little dust closer to the star than 40 AU (Jura et al. 1993). The existence of one or more planets was suggested as one of the possible causes of this central hole in the dust distribution (Jura et al. 1995; Jura et al. 1998). J98 and K98 determined that the disk's diameter is ~3", which corresponds to 200 AU at the star's distance of 67 pc, and they confirmed directly the existence of the central hole. An especially significant aspect of the discovery is that, based on isochrone fitting and the strong lithium absorption line, the presumably coeval companion HR 4796B, located 7.7" away at PA=226° (Jura et al. 1993), is an M2-M3 pre-main sequence star with an age tightly constrained to be $(8 \pm 3) \times 10^6$ yr (Stauffer et al. 1995; Jura et al. 1998). Observational and theoretical studies of our solar system, including chronological data for meteorites, suggest that ~$10^7$ yr is a reasonable estimate of the timescale for the duration of the solar nebula and the growth of the planets to nearly their final masses (see reviews by Lissauer 1993 and Podosek & Cassen 1994). Thus, the constraints on the age of HR 4796A place it within a critical stage of planetary disk evolution.

Our new images clarify the structure of the central clearing zone, or hole, which may be indirect evidence for a planet orbiting HR 4796A (see, e.g., Roques et al. 1994). They may also reveal an asymmetry in the disk, possibly confirmed by the NICMOS images (Schneider et al. 1999), that could reflect the influence of the stellar companion or a hypothetical planet. Finally, these multi-wavelength data provide a reasonable basis for estimating a characteristic size for the mid-infrared-emitting dust particles, an estimate with implications for our understanding of the disk evolution. Our results and discussion complement those in a companion paper by Wyatt et al. (2000, hereafter W2000).

## 2. MID-INFRARED IMAGING WITH OSCIR AT KECK II

We imaged HR 4796A in the N (10.8 µm) and IHW18 (18.2 µm) bands at the Keck II Telescope in 1998 May 7-12 with OSCIR, the University of Florida mid-infrared imager/spectrometer. OSCIR contains a 128 × 128-pixel, Si:As Blocked Impurity Band (BIB) array developed by Boeing. The plate scale was 0.062 arcsec/pixel with the f/40 chopping secondary at the Keck II visitor bent-cassegrain port, which gave an array field of view of 7.9" × 7.9". We used a guide star and the off-axis autoguider, which ensured excellent tracking and image stability, and the standard chop/nod technique, with a chop frequency of 4 Hz and a throw of 8" in declination. The stars $\alpha$ Sco, $\gamma$ Aql, and $\delta$ Vir were used for flux calibration. The seeing was generally very good for our observations. Typical values of the full width at half-maximum intensity (FWHM) of the observed point spread functions (PSFs) were 0.32" at 10.8 µm and 0.45" at 18.2 µm, compared to the diffraction-limited values ($\lambda/D$) of 0.22" and 0.37", respectively. Quadratic subtraction of these implies 10-20 µm "seeing" of $\sim$0.2"-0.3" (which may include other contributions, e.g., from autoguider jitter). The diffraction limits at 10 and 18 µm correspond to four and six pixels, respectively, and thus OSCIR spatially samples the diffraction-limited PSF more finely than conventional Nyquist sampling. We encountered light cirrus throughout much of our Keck run. By comparing observations within and among the nights, we believe that we have been able generally to remove the data for which the photometry is most uncertain. However, because of the cirrus, some unquantifiable uncertainty remains in the photometry. Key properties of the observations, including flux densities and noise estimates, are listed in Table 1. The images are presented as contour plots in Figures 1a and 1b. Figure 1c shows the unsmoothed 18.2 µm image, which indicates the quality of the original data.

## 3. DISK MORPHOLOGY IN THE THERMAL INFRARED

The discovery images clearly showed the disk around HR 4796A to be extended at 18 µm, but both teams only marginally resolved it at 10 µm (J98; K98). Figure 1 demonstrates that the emission ridge corresponding to the edge-on disk is well resolved at both 10 and 18 µm in these new images. For comparison, the PSFs determined from nearby stars are presented in Figure 2 on both logarithmic and linear brightness scales. At 10 µm, the disk is evident as very low-level wings protruding from the strong stellar point source (which, for convenient display, is therefore plotted logarithmically in Figure 1a). The ridge is much more prominent at 18 µm (plotted linearly in Figure 1b). The line connecting the two outer peaks is oriented along PA=26°±2°, and the maximum extent of the lowest 18 µm contour displayed in Figure 1 is about 3.1", or 208 AU. The maximum extent at 10 µm is about 2.8", or 85–90% that of the 18 µm emission, and thus the 10 µm and 18 µm sources are roughly equal in size along the ridge. We also resolved the 18 µm disk along the minor axis, as indicated by the scans through the lobes and a comparison star shown in Figure 3.

Three peaks of comparable brightness are embedded in the 18 µm ridge. Each of the two outer 18 µm peaks, which we refer to as the lobes, is separated from the central peak by 0.8" (54 AU). The brightest region within the central 18 µm contour is indicated in Figure 1b by a filled circle. That point is slightly offset from the centroid of the highest central contour; it lies on the line connecting the two lobes and halfway between them. By imaging sequentially at the two wavelengths with the Keck II visitor-port autoguider in operation, which facilitated accurate registration, we determined that the central 18 µm peak coincides with the 10 µm peak to within 0.3". As described below, most of the 10 µm emission is direct starlight, and therefore the coincidence between the central 10 and 18 µm peaks implies that the central 18 µm peak also coincides with the star, which we therefore assume is coincident with the filled circle in Figure 1b. In the 10 µm contour plot in Figure 1a we indicate (crosses) the location of the peaks of the 18 µm lobes. The 18 µm lobes clearly coincide with the 10 µm wings, and therefore much, possibly most, of the 10 µm-emitting dust must be mixed with, or identical to, the 18 µm-emitting dust. The 18 µm brightness at the lobes is comparable to that at the star, whereas their 10 µm brightness is only 4–5% that at the star. The disk morphology that we see at 18 µm is consistent with that seen by K98 in their Keck images. K98 did not resolve the disk at 10 µm.

The mid-infrared radiation from HR 4796A originates both in the stellar photosphere and in dust that is heated by the starlight. The stellar photospheric flux densities at 10 and 18 µm are about 80% and 5%, respectively, of the total observed flux densities. Although the 18 µm stellar flux is a small fraction of the total, it contributes a large fraction of the emission of the central peak itself. To determine the contribution of the stellar photosphere to the image, we subtracted the stellar PSF from the image of HR 4796A after registering the peaks and normalizing the comparison star's flux to that inferred for HR 4796A (Table 1). The PSF star was located near HR 4796A and observed just before and just after it. Due to the hexagonal shape of the Keck primary and secondary mirrors, the PSF is hexagonally shaped, so that, for the most accurate point-source subtraction or deconvolution, one cannot assume circular symmetry of the PSF. During the 18 µm observations of HR 4796A, the PSF rotated about 23° relative to the imager field of view (a rotation inherent in the alt-az mount). From the observed PSFs we constructed a time-averaged PSF appropriate to the rotational range of the PSF during our observations of HR 4796A.

The slightly smoothed (three pixels) original and star-subtracted 18 µm images are shown in Figures 4a and 4b, respectively. The star was assumed to be coincident with the brightest part (filled circle, Fig. 1b) of the central maximum. Subtraction of the stellar flux removes the central peak, leaving a hole at that position. Because of the apparent absence in HR 4796 of significant dust hotter than about 110 K, Jura et al. (1995) concluded that the disk has a roughly 40 AU-radius central hole in it. This conclusion was subsequently confirmed by the first direct images of the disk at ~20 µm (J98; K98) and, more recently, by NICMOS images at 1-2 µm (Schneider et al. 1999; see below). We see the hole directly in our 18 µm image (Figures 1b & 4a) as a depression between the central peak and each of the two lobes; it is obvious in the image with the starlight subtracted (Figure 4b). Some tenuous emission may also be present in the inner region, i.e., the region within ~0.5" of the star. The appearance of this residual emission depends on how we characterize and position the subtracted stellar PSF. In the Appendix we address briefly some aspects of how the computed residual is effected by slight errors in the position and width of the PSF used to subtract HR 4796A's starlight.

We also subtracted the stellar photospheric 10 µm flux from the 10 µm image. The central hole is evident in that subtracted image, but the disk is so much fainter than the star that the subtraction procedure is very uncertain. Therefore, we do not have confidence in the accuracy of this result insofar as it bears on the structure of the inner part of the emitting region. However, those 10 µm results are useful in the context of estimating the characteristic size of the mid-infrared -emitting particles (§6).

There is some evidence that the lobes are not of equal maximum brightness: the NE lobe is 5.9 ±3.2% brighter than the SW lobe, where this asymmetry corresponds to the difference in the lobe peak brightnesses divided by their mean. This value for the asymmetry includes a 0.8% correction for an asymmetry in the PSF. The formal statistical significance of the brightness asymmetry of the lobes is therefore 1.8 σ. The brightness asymmetry itself was determined by fitting a polynomial to a 10-pixel×10-pixel (0.6" ×0.6") region approximately centered on each peak. The uncertainty associated with this asymmetry was established by convolving the best-fit model (§4) with a gaussian approximation to the PSF, applying gaussian noise like that observed across the array, and fitting the polynomial as mentioned above. After repeating this procedure 50,000 times, we computed the rms deviation, which constituted the formal uncertainty. This asymmetry may also be apparent in the 10 µm image (particularly in the scan shown in Figure 8b, discussed below) and in the NICMOS image (Schneider et al. 1999). The asymmetry is discussed in a much broader context in the companion paper by W2000. We briefly comment on it below.

### 4. MODELING THE BRIGHTNESS DISTRIBUTION

J98 and K98 modeled their ~20 µm images of HR 4796A to infer structural properties of the disk. Both models assumed optically thin, passive disks in which the dust particles have an infrared emission efficiency varying with wavelength as $\lambda^{-1}$. J98 found that an edge-on dusty disk with an inner hole radius somewhere in the range 40–80 AU provides an adequate fit to the spectral energy distribution and the image. The higher resolution Keck image obtained by K98 more tightly constrains their thin-disk model, from which they infer an inner hole radius of 55 ±15 AU and a disk inclination of 18° (+9°,-6°) from edge on.

In the companion paper, W2000 present results of modeling the 18 µm brightness distribution of HR 4796A. Since it is relevant to our estimate of the grain size and to the comparison of our results to the HST NICMOS observations, we summarize the model here. W2000 consider the dynamical interaction of a circumstellar disk and a stellar or planetary companion, a study initially motivated by the desire to understand how such interactions generate asymmetries and other structures in circumstellar disks. A distribution of orbiting dust particles is gravitationally perturbed by a companion, and W2000 examine the resultant orbital and density distributions for dust grains of a given size and composition. W2000 ignore the effects of radiation pressure, assuming, in effect, that large, dynamically stable dust grains, with $\beta \approx 0$, orbit HR 4796A; $\beta$ is the ratio of the radiation pressure force to the gravitational force on a particle (called $\delta$, the overpressure ratio, by Backman & Paresce 1993). However, at least in the context of Mie theory, we know that these large particles are not the ones that we observe in the mid-infrared. As discussed in §6, the mid-infrared–emitting particles, if Mie spheres of astronomical silicates (Draine & Lee 1984), are ~2-3 µm in diameter, with the corresponding value $\beta \approx 2$. Radiation pressure will drive particles with such a high value of $\beta$ out of the system on hyperbolic orbits in a few hundred years. Based on this negligible lifetime and on estimates of the particle number density and collision rates in the disk, W2000 (§6) conclude that these grains must be collision fragments of much larger particles. The spatial distribution of larger particles in the HR 4796A disk is unknown, so, as an alternative, W2000 assume that larger particles are distributed essentially like that of the smaller, 2-3 µm-size, Mie grains emitting the observed mid-infrared radiation. By accounting for the temperature dependence of the dust as a function of distance from the star, they determine the spatial distribution of the mid-infrared-emitting grains from our 18 µm image.

The model disk is not flat, but flared, with a distribution of orbital proper inclinations and proper eccentricities identical to those of the main-belt asteroids in our own solar system. The distribution of semi-major axes for the particles is assumed to follow the relation $n(a) \propto a^{\gamma}$, where $n(a)$ is the number of orbits per unit semi-major axis. For particle orbits that are circular, which is approximately true for the model orbits, the surface number density as a function of distance $r$ from the star is therefore proportional to $r^{\gamma-1}$. The largest value, $a_{max}$, used for the semi-major axis of the dust orbits is 130 AU, although the results of the modeling are not sensitive to this value, since relatively little flux arises near the outer edge of the disk. Key model parameters that are varied to match the observations are: (1) the exponent $\gamma$; (2) the minimum value, $a_{min}$, of the dust semi-major axis, which controls the diameter of the central clearing; (3) the normalization of the density distribution; and (4) the inclination of the disk symmetry plane to the line of sight. In addition, observed asymmetries in the brightness distribution are modeled by adjusting the common forced eccentricity and the direction of the forced pericenter of the disk particle orbits.

Figure 4c shows the model brightness distribution at 18 µm that is the best fit to the star-subtracted image in Figure 4b. The residual (observed minus model) is shown in Figure 4d. The model fits the lobes well, with nearly all of the residual flux located within 0.4" of the star. We are not certain of the nature of the residual flux. Some of it may represent real flux from HR 4796A's inner disk, but some of it may result from imperfect subtraction of the stellar photospheric flux stemming from variations in the seeing not manifested in the PSF-star images (which were of shorter duration). We briefly address this issue in the Appendix.

The model disk inclination is $13° \pm 1°$ from edge-on. A flat disk would be modeled as more face-on than this, since the disk flaring that we assume in the model accounts for some of the resolved disk extension that we see along the minor axis (Figure 3). For the distribution of semi-major axes, $\gamma = -2 \pm 1$. The resultant model distribution of surface number density (i.e., integrated normal to the disk) of 2.5 μm grains is shown in Figure 5 in the form of a radial cut through the azimuthally averaged disk (solid curve). The surface density is relatively sharply peaked near 70 AU. Exterior to this position, the surface density decreases with distance approximately as $r^{-3}$. Interior to this position, the surface density drops steeply by a factor of two between 70 and 60 AU, falling to zero by 45 AU. The model fit implies $a_{min} = 62 \pm 2$ AU, but because the dust particle orbits are assumed to be non-circular, the hole is not sharply defined in the models; i.e., the orbital eccentricity results in a small range in a particle's radial location, which softens the hole's edge. Flat-disk models based on the disk-discovery images indicate comparable values for the hole radius (J98; K98), but the dynamical models considered here provide an *a priori* basis for understanding why a hole should have an indistinct, or soft, edge.

How empty is the hole in the disk? We can estimate the optical depth in the hole by comparing the brightness of the residual emission there (assumed to represent an upper limit to emission from dust) to the observed brightness of the lobes. Consider first the optical depth in the disk lobes. The SW lobe is 54 AU in projection from the star. The model temperature there is 113 K (compared to the blackbody temperature of 81 K) and, ignoring projection effects, the corresponding edge-on optical depth at 18 μm is ~0.002. The residual emission (observed minus model; Fig. 4d) at 20 AU (0.3") from the star is typically 0.2 that of the lobes. The model temperature at 20 AU is 163 K, and thus the 18 μm edge-on optical depth there is $\sim 5 \times 10^{-5}$, which is about 3% that of the lobes. Assuming that the ratio of the face-on optical depth to the edge-on optical depth in the hole is comparable to that ratio at the lobes, then the average optical depth of 18 μm-emitting dust in the hole appears to be about 3% that in the main part of the disk at 60-100 AU from the star. If any of the residual emission results from poor PSF subtraction and is not intrinsic to the source, the optical depth of the dust in the hole could be even less than this. The central cleared region appears to be relatively very empty.

There has been speculation about the causes of holes in the centers of circumstellar disks (Lagage & Pantin 1994; Jura et al. 1998; Strom, Edwards, & Skrutskie 1993; Roques et al. 1994; J98; K98), and the disk in HR 4796A is likely to become a strong focus for these discussions (e.g., Jura et al. 1995; W2000). We only wish to emphasize here that the possible disk brightness asymmetry is an interesting and perhaps unexpected manifestation of the central hole. The brightness asymmetry in the model results from the gravitational perturbation of the disk particle orbits by the low-mass companion HR 4796B even if the companion's orbit is only moderately eccentric (W2000). To first order, the effect of the perturbation is to shift the center of symmetry of the circular disk from HR 4796A towards the companion's apocenter, where the companion spends most of its time; i.e., the long-term average of the perturbation on the disk particles is directed towards the companion's apocenter. The side of the disk, and therefore the hole edge, that is closest to the companion's pericenter is shifted closer to the primary star and is therefore warmer ("pericenter glow"). The pericenter glow is a direct result of there being a hole in the disk. If this model is valid and these perturbations do arise from HR 4796B, then HR 4796B's pericenter must be located NE of HR 4796A, and HR 4796B is currently located near its apocenter SW of HR 4796A. The scale of the shift of the disk's center of symmetry is indicated by comparing the dashed and dotted curves in Figure 5, which shows the density distribution through the disk towards and away from the stellar companion's pericenter. The disk is shifted only a few AU, which, while small, leads to an arguably measurable brightness asymmetry. The perturber responsible for the shift of the disk need not be exterior to the disk. A star or planet within the hole could cause the same effect (W2000); however, based on near-infrared speckle observations (Jura et al. 1995), the mass of any stellar companion located between 11 and 120 AU from HR 4796A must be less than $0.13 M_\odot$. The fact that we may be detecting a brightness asymmetry resulting from a few-AU shift in the disk centroid implies that the detection of significant dynamical effects on circumstellar disks by stellar or planetary companions is now within reach. The HST/NICMOS images presented by Schneider et al. (1999) may confirm this pericenter glow (§5).

## 5. COMPARISON OF MID-INFRARED AND HST/NICMOS IMAGES

Schneider et al. (1999) have imaged the HR 4796A disk at 1.1 and 1.6 µm with NICMOS. They used a 0.6"-diameter coronographic spot, but, due to a variety of instrumental effects, the data in a ~1.3"-diameter region are unusable. Their images show clearly what appears to be a relatively narrow ring inclined to the line of sight. Assuming that the ring is intrinsically circular, they determine a major-axis position angle of $26.8° \pm 0.6°$ and an inclination to the line of sight of $73.1° \pm 1.2°$, in agreement with our estimates (W2000; §§3 & 4) and K98's. The mean radius of the ring is 70 AU, and, after quadratic subtraction of the PSF, the full width of the annulus is estimated to be 17 AU (Schneider et al. 1999). The relatively sharp maximum in the density near 70 AU determined from our modeling of the OSCIR observations agrees precisely with the value derived by Schneider et al. (1999) for the mean radius of the ring based on the NICMOS observations. We now consider a more detailed comparison.

In Figure 6 we show our 10 and 18 µm contours overlaid on the 1.1 µm NICMOS image. Keeping in mind the differences in the resolutions of the images (FWHM: 0.12" at 1.1 µm; 0.32" at 10.8 µm; 0.45" at 18.2 µm), we see that the overall coincidence of the mid-infrared and near-infrared emission is excellent. However, the 18 µm lobes are located noticeably closer to the star than are the bright 1.1 µm peaks. The peaks of the 18 µm lobes are separated by 1.73", compared to 2.10" for the peak-intensity positions at 1.1 µm (Schneider et al. 1999). To assess the nature of this offset, we considered a perfectly thin ring (annulus) of dust with uniform number density and the temperature distribution derived in the W2000 models. We assumed a mean annular radius of 70 AU and an annular width of 17 AU and computed the 18.2 µm brightness at each point in the ring, which we convolved with the observed 18.2 µm PSF. In Figure 7 we compare the brightness distribution along the major axis expected for this convolved model to the observed brightness distribution.

This comparison implies that ~60% of the 0.37" difference in the separation of the peaks at 1.1 and 18.2 µm results from the lower resolution of the 18 µm image. Figure 7 indicates that the uniform-annulus model results in an 18 µm distribution with lobes displaced to larger radii and appearing somewhat narrower than actually observed. The discrepancy in slope is primarily on the insides of the lobes. The observed outer slope of the NE lobe is well matched by the model, while the match of the SW lobe is noticeably poorer. We do not think that these differences between the NICMOS-based model and the 18 µm observations result from inaccurate subtraction of the stellar PSF from the 18 µm image. Figure 7 gives some indication of the small dependence of the shape inferred for the inside of the disk lobes on the assumed shape of the PSF; we show the flux distributions resulting from use of the nominal, observed PSF (solid curve) and the same PSF with an additional 0.35" of seeing (dashed line; see Appendix). Rather, the difference in the observations and the brightness expected from the model may result from the annulus having a somewhat soft inner edge. Primarily because of the lower resolution of the 18 µm observations, a more gradual density falloff on the inside of the annulus will both shift the observed peaks inward and lower the inner slope of the lobes. The excellent agreement in the location of the peak density in the disk, as determined from models of the OSCIR images, and the average radius of the annulus determined from models of the HST/NICMOS images, suggests that it is this inner density rolloff, or a similar refinement such as excess emission in the hole, that accounts for the relative locations of the observed emission peaks in the two datasets. A slower density falloff on the *outside* of the annulus would elevate the 18 µm wings, although we see this only on the SW side of the disk. As shown by W2000 and discussed in the previous section, one may account for a soft density falloff on the inside of the annulus by the kinds of dynamical properties anticipated (and assumed in the W2000 models) for the orbiting dust particles.

Figure 6 shows that at 1.1 µm the NE side of the annulus is noticeably brighter than the SW side. This asymmetry is not evident in the 1.6 µm NICMOS image. This asymmetry may support our, albeit marginal, detection of the pericenter glow at 18 µm. However, one is seeing emission from dust particles at 18 µm, but scattering from dust (and probably a different population) at 1-2 µm. We speculate that the mass shift that may result in the enhanced lobe brightness at 18 µm could also enhance the scattering of light there. Detailed models that incorporate realistic scattering functions must be carried out to explore this interesting possibility.

## 6. THE MID-INFRARED –EMITTING DUST

### *6.1 The Characteristic Particle Size*

Circumstellar dust spans a range of compositions and sizes. However, for a given composition, one can derive from observations at two wavelengths a unique particle size that is arguably characteristic of the dust responsible for that emission. The basis for this estimate is that the emission efficiency, and therefore the temperature, of a dust particle depends on its size. We made an initial estimate of the size of dust particles in HR 4796A by taking the ratio of the 10 and 18 µm fluxes in the lobes (after correcting for the different PSFs at the two different wavelengths) and comparing it with that expected from Mie silicate spheres of different sizes, which have different emission efficiencies and consequently different temperatures. The formal value of this initially estimated diameter was 2.5 µm. (Mie silicate grains of this diameter have infrared emission efficiencies varying approximately as $\lambda^{-1.9}$.) This initial estimate was then used to do the modeling described in §4, which was repeated using 2 and 3 µm-diameter particles. In Figure 8a we show the resultant scan along the major axis of the 18 µm image from which the stellar photospheric emission (using the nominal PSF; see Appendix) has been subtracted. The model provides a good fit to the 18 µm lobe structure for a range of assumed dust sizes. However, Figure 8b shows how the 10 µm brightness constrains the diameter to 2.5 µm, consistent with our original estimate. We conclude that, under the assumption that the particles we see in the lobes are astronomical silicate Mie spheres, 2-3 µm-diameter grains provide a reasonable fit to the lobe mid-infrared colors. Of course, based on studies of interplanetary dust particles in our solar system (e.g., Gustafson 1994), it is evident that the dust particles in HR 4796A's disk are unlikely to be simple spheres, the implications of which we are in the process of considering.

*6.2 Particle Evolution*

We can consider the effects of the stellar radiation on 2.5 µm-size grains by estimating the value of the parameter β, the ratio of the radiation pressure force to the gravitational force on a particle. As summarized in Artymowicz (1988) and elsewhere, $\beta \propto \langle Q_{pr} \rangle (L_*/M_*)/R\rho$, where R is the particle radius, ρ is its density, $\langle Q_{pr} \rangle$ is the radiation pressure efficiency averaged across the stellar spectral energy distribution, and $L_*/M_*$ is the ratio of the stellar luminosity and mass. W2000 have computed the value of β for 2.5 µm-diameter silicate spheres near HR 4796A; they assumed $L_*=21L_\odot$, $M_*=2.5M_\odot$, and ρ=2.5 g cm$^{-3}$, and they compute $\langle Q_{pr} \rangle$ using a stellar spectral energy distribution for the A0V star Vega (M. Cohen, private communication). They derive the value β=1.7 for the grains that characterize the mid-infrared–emitting particles in HR 4796A's disk. If one takes into account the differences in spectral type and assumed grain density, this value agrees reasonably well with the one that can be derived from the results in Artymowicz (1988, see his Figure 1), who computed β for the star β Pic and for a variety of grain materials including the astronomical silicates of Draine & Lee (1984). Dust grains that are created in collisions, and for which β is greater than ~0.5, have hyperbolic orbits; they are blown beyond ~100 AU and out of the system on essentially free-fall time scales, on the order of a few hundred years. Thus, the 2-3 µm grains we see orbiting HR 4796A are not primordial; one expects very few of them in this disk (see also Jura et al. 1995). The detailed analysis by W2000 (see also Artymowicz 1997) implies that the collisional lifetime of the dust in HR 4796A is roughly $10^4$ yr across the disk. The mid-infrared-emitting particles are probably relatively recent fragments of collisions that dominate both the creation and destruction of dust there.

## 7. SUMMARY AND CONCLUSIONS

1. Our new 10 and 18 µm images of the circumstellar disk around the A0 Vega-type star HR 4796A images show extended, well resolved emission at 10 µm in addition to the previously discovered 18 µm emission (J98; K98). The resolution of both images is close to the Keck diffraction limit. The size of the 10.8 µm-emitting region is comparable to that at 18 µm, implying that most of the 10 µm-emitting particles are mixed with, or identical to, those emitting at 18 µm.

2. Modeling of the disk brightness distribution (W2000) implies that a pronounced maximum in the surface density occurs at approximately 70 AU from the star. This value equals the mean annular radius deduced by Schneider et al. (1999) from their HST/NICMOS images. The best-fit model surface density exterior to 70 AU falls steeply with distance as $r^{-3}$. Interior to 70 AU, the density drops steeply by a factor of two between 70 and 60 AU, falling to zero by 45 AU, which corresponds to the edge of the previously discovered central hole; in the context of the dynamical models, this "soft" edge for the central hole results from the fact that the dust particle orbits are non-circular.

3. Based on the images, the optical depth in the central clearing zone, or hole, is ~3%, or less, that in the main part of the disk. Thus, the hole is indeed relatively empty.

4. We present tentative evidence (1.8 $\sigma$) that the 18 µm lobes are of unequal brightness. Detailed modeling of the 18 µm brightness distribution based on dynamical considerations and presented in a companion paper (W2000) indicates that such a brightness asymmetry could result from gravitational perturbation of the disk particle orbits by the low-mass stellar companion or a planet. The perturbation shifts the disk center of symmetry away from the companion's pericenter; to first order, this shift corresponds to a spatial shift of the center of the circularly symmetric disk. This shift brings the inside edge of the annulus closer to HR 4796A, which makes that edge warmer and more infrared-emitting ("pericenter glow") than the other side of the annulus. This measurable brightness asymmetry corresponds to a spatial shift of the disk of only a few AU, and so the detection of even fairly complex dynamical effects of planets on disks appears to be within reach. The recent 1.1 µm NICMOS image (Schneider et al. 1999) may show evidence in support of the pericenter glow.

5. We compare our 10 and 18 µm images with recent NICMOS images of HR 4796A at 1.1 and 1.6 µm that show what appears to be a relatively narrow ring, or annulus, inclined to the line of sight and with a mean radius of 70 AU and an annular width of ~17 AU. The overall coincidence of the near-infrared and mid-infrared structure is excellent. However, even after accounting for differences in angular resolution, we find that the 18 µm distribution expected from the NICMOS-based model differs somewhat from the observed distribution. These differences suggest that the annulus has a "soft" inner edge, which may correspond to the density falloff expected from considerations of particle dynamics.

6. Using the 10 and 18 µm images, we estimate that the mid-infrared-emitting dust particles in the main part of the disk have a characteristic diameter of ~2-3 µm. This estimate is based on the assumption that the particles are Mie spheres composed of astronomical silicates. The corresponding value of the parameter β, the ratio of the radiation-pressure force to gravitational force on the particle, is 1.7. Particles with β >0.5 are expected to be quickly blown out of the system. We conclude that the mid-infrared–emitting particles in the HR 4796A disk are not primordial, but rather they are relatively recent and transient collision fragments.

## ACKNOWLEDGEMENTS

We wish to acknowledge Kevin Hanna for his development of the superb OSCIR electronics; Kevin Hanna, Jeff Julian, and Chris Singer for their assistance in preparing for, and carrying out, the Keck observing run; Martin Cohen for providing us with the Vega spectrum; and the Keck Observatory personnel for their outstanding support of OSCIR as a Keck II visitor instrument. Use of OSCIR at Keck was supported in part by NSF and NASA grants to the University of Florida and the Pennsylvania State University at Erie.

# APPENDIX

We pointed out in §3 that some of the inner tenuous emission (within 0.5" of the star) apparent in the star-subtracted 18 μm image of HR 4796A (Figure 4b) and as residual emission in Figure 4d may be an artifact resulting from the use of an unsuitable PSF. This mismatch could occur if there are low-frequency seeing components sampled during the ~1 h of chopped integration time on HR 4796A (Table 1) that are not sampled during the much shorter (~1 m) integration time on the nearby star used to determine the PSF. (Given the limited amount of available observing time, it is unrealistic to observe the PSF star for the same amount of time that we observe the program objects.) Here we briefly examine this issue.

Figure 9a presents the star-subtracted 18 μm image presented previously in §3 (Figure 4b). That image resulted from the subtraction of HR 4796A's direct starlight (photospheric emission) using the observed stellar PSF (referred to as the nominal PSF) and assuming that the stellar peak is coincident with the location of maximum 18 μm brightness (filled circle, Figure 1). Figure 9b shows the result of instead subtracting a somewhat broader PSF obtained by convolving the nominal PSF with a gaussian profile with a FWHM of 0.35". This might represent some additional seeing that could have affected the image of HR 4796A but not the observed PSF star image. In Figures 9c and 9d, we repeat this exercise but with the star assumed to be at the centroid of the highest central contour indicated in Figure 1b, i.e., at 0.1" WNW of the filled circle in Figure 1b. The intent of this exercise is to illustrate the sensitivity of the residual inner structure to the assumed stellar location.

We see that even this modest amount of extra "seeing" results in a star-subtracted image of HR 4796A that is noticeably less structured at the disk center. This comparison is also shown in the scans in Figure 7. Likewise, slight errors in positioning the star result in some differences in the inferred inner-disk structure. However, although the emission structure differs somewhat, the integrated residual flux is virtually unchanged, since the stellar flux is the same in both cases. As we have noted, the light cirrus evident throughout much of our Keck run may have resulted in poor photometric conditions, but we do not think this is the source of the residuals: the photometry is generally consistent to within ~20%, and even an error this large in the PSF flux normalization would only marginally effect the residual.

We conclude that the detailed structure of the inner disk (within 0.5" of the star), as inferred from the star-subtracted image or the resultant image of the residual emission (e.g., Figure 4d), must be viewed with caution. However, the integrated residual flux is more robust and may be real. If so, it may correspond to low-level dust emission from within the central clearing.

# FIGURE CAPTIONS

**Figure 1.** Images of HR 4796A made with OSCIR. North is up, and east is to the left. The images from which the two contour plots were constructed were smoothed with a 3-pixel (FWHM), or 0.19 arcsec, gaussian. For both contour plots, the lowest contour level is about four times the smoothed noise. **(a)** 10.8 μm image with contours (mJy per arcsec$^2$) spaced logarithmically at 6.4, 9.9, 15, 23, 36, 55, 85, 132, 203, and 312. The filled circle is the location of the peak brightness, with the value 480 mJy per arcsec$^2$. The crosses indicate the positions of the 18.2 μm lobes; **(b)** 18.2 μm image with contours (mJy per arcsec$^2$) spaced linearly at 58, 93, 128, 163, 198, 234, 269, 304, 339, and 374. The filled circle indicates the presumed position of the star, where the brightness is 362 mJy per arcsec$^2$; **(c)** Unsmoothed version, with linear brightness levels, of the 18.2 μm image.

**Figure 2.** Plots of the point spread functions (PSFs) at 10.8 and 18.2 μm based on observations of comparison stars. North is up, and east is to the left. All plots are normalized to a maximum value of 100 units located at the filled circle. **(a)** Linear 10.8 μm levels: 1, 12, 23, 34, 45, 56, 67, 78, and 89; **(b)** Logarithmic 10.8 μm levels: 1.3, 1.9, 2.8, 4.2, 6.2, 9.2, 13.7, 20.2, 29.9, 44.3, and 65.5; **(c)** Linear 18.2 μm levels: 8.0, 18.2, 28.4, 38.6, 48.8, 59.0, 69.2, 79.5, and 89.7; **(d)** Logarithmic 18.2 μm levels: 8.0, 10.6, 13.9, 18.4, 24.3, 32.1, 42.3, 55.9, and 73.7.

**Figure 3.** Comparison of minor-axis scans through the 18.2 μm lobes and a comparison star, which shows that the lobes are resolved along the minor axis of the disk. The northwest side of the scan is to the right (positive offsets).

**Figure 4.** Comparison of the 18.2 μm images of HR 4796A with the model brightness distribution. East is up and north is to the left. Overlaid contour levels (mJy per arcsec$^2$) are linearly spaced. **(a)** Observed image smoothed with 3-pixel gaussian, with contour levels same as in Fig. 1b; **(b)** Observed image after subtraction of stellar photospheric emission. Contour levels: 58, 93, 128, 163, 198, 234, 269, 304, 339, and 374; **(c)** Model brightness distribution; **(d)** Residual emission, corresponding to the difference between the star-subtracted and model brightness distributions. Contour levels: 20, 40, and 60.

**Figure 5.** Plots of the surface number density of the 2.5 μm dust grains in the disk derived from modeling the 18 μm brightness distribution. The solid curve is the azimuthal average of the surface number density, the dotted line and the dashed line indicate the density through the disk in the direction of and away from, respectively, the stellar companion's epicenter. The relative

displacement of the dotted and dashed lines indicates the scale of the displacement of the disk center from the star.

**Figure 6.** Overlay of **(a)** 10.8 and **(b)** 18.2 μm contours on the 1.1 μm HST/NICMOS image of HR 4796A disk from Schneider et al. (1999). The angular resolutions (FWHM) of the images are: 0.12" at 1.1 μm; 0.32" at 10.8 μm; 0.45" at 18.2 μm. In both images, east is approximately down, and north is to the left (see Schneider et al. 1999). In this orientation, the companion star HR 4796B is located to the upper right.

**Figure 7.** Comparison of major-axis cut through the lobes of our 18 μm star-subtracted image (solid curve) to that (dotted curve) expected from the simplest disk model inferred by Schneider et al. (1999) from their NICMOS image. The left side of the scan is northeast of the star. The model disk is a thin annulus with a mean radius of 70 AU, a width of 17 AU, a uniform density, and the temperature falloff with radius determined by W2000. The Schneider et al. model has been convolved with the observed 18 μm PSF. The dashed curve shows the result of subtracting from the observations a stellar PSF that has an additional 0.35" seeing, as discussed in the Appendix.

**Figure 8.** Comparison of **(a)** the star-subtracted 18.2 μm and **(b)** 10.8 μm scans, indicated as solid lines, along the disk major axis of the HR 4796A disk. The left side of each scan is northeast of the star. In the 18 μm scan, the features within ~0.3" of the stellar position are probably artifacts resulting from imperfect PSF subtraction. Only the lobes of the 10 μm scan are shown (see text). The 18 μm brightness distribution has been modeled, as described in the text, using spherical Mie silicate particles that are 2, 2.5, and 3 μm in diameter. The corresponding scan of the 10.8 μm flux distribution expected from a disk composed of each of these particles indicates that the 10.8 μm brightness in the lobes is best fit by 2.5 μm-diameter particles. Thus, in the context of Mie theory, the grains are constrained to be a few microns in diameter.

**Figure 9.** Comparison of the 18 μm disk emission resulting from the subtraction from the observed image (Figure 1b) of a modest range of stellar point spread functions normalized to the expected flux of the star HR 4796A. North is up, and east is to the left. As discussed in the Appendix, the intent of this exercise is to illustrate the effect on the inferred inner disk structure of the use of an inaccurate, or slightly mis-positioned, PSF. **(a)** Nominal observed PSF (FWHM=0.45") subtracted at brightest spot (filled circle in Figure 1b); **(b)** Seeing-convolved PSF (seeing FWHM=0.35") subtracted at brightest spot; **(c)** Nominal observed PSF subtracted at centroid of highest central contour; **(d)** Seeing-convolved PSF subtracted at centroid of highest central contour.

**Table 1: OSCIR Observations of HR 4796A**

| Filters | N | IHW18 |
|---|---|---|
| $\lambda_o$ (μm) | 10.8 | 18.2 |
| $\Delta\lambda$ (μm) | 5.3 | 1.7 |
| Int. Time (s) [a] | 2784 | 3480 |
| Total $F_\nu$ (Jy) [b] | 0.188 ±0.047 | 0.905 ±0.130 |
| 1σ Noise/Pixel (mJy) | 0.014 | 0.15 |
| Stellar $F_\nu$ (Jy) [c] | 0.150 | 0.048 |
| αSco $F_\nu$ (Jy) | --- | 817 |
| γAql $F_\nu$ (Jy) | --- | 26 |
| δVir $F_\nu$ (Jy) | 142 | --- |

[a] Integration times correspond to both halves of the chopper cycle. They are the sum of the times for which the source itself and the reference sky position were projected onto the detector.

[b] The uncertainties for the 10.8 and 18.2 μm flux densities, which are based on the reproducibility of standard-star fluxes, are principally due to the presence of light cirrus clouds. This problem was more severe for the 10.8 μm than for the 18.2 μm image.

[c] The 10.8 μm photospheric flux density was derived by Jura et al. (1998) by extrapolating the K-band (2.2 μm) flux density to 10.8 μm using the Kurucz (1979)